\documentclass{article}
\usepackage{amsmath}
\usepackage{graphicx}

\def\Mb{\kern 2pt\mathchoice
            {             \vbox{\hrule width10pt height 0.4pt depth 0pt
                 \kern 1.2pt\hbox{\kern -2pt$\displaystyle M$}}}
            {                 \vbox{\hrule width10pt height 0.4pt depth 0pt
                 \kern 1.2pt\hbox{\kern -2pt$\textstyle M$}}}
            {\vbox{\hrule width6pt height 0.4pt depth 0pt
                 \kern 1.0pt\hbox{\kern -2pt$\scriptstyle M$}}}
            {                 \vbox{\hrule width5pt height 0.4pt depth 0pt
                 \kern 0.8pt\hbox{\kern -2pt$\scriptscriptstyle M$}}}}
\def\Sb{\kern 2pt\mathchoice
            {                 \vbox{\hrule width6pt height 0.4pt depth 0pt
                 \kern 1.2pt\hbox{\kern -2pt$\displaystyle S$}}}
            {                 \vbox{\hrule width6pt height 0.4pt depth 0pt
                 \kern 1.2pt\hbox{\kern -2pt$\textstyle S$}}}
            {                 \vbox{\hrule width3.5pt height 0.4pt depth 0pt
                 \kern 1.0pt\hbox{\kern -2pt$\scriptstyle S$}}}
            {                 \vbox{\hrule width3pt height 0.4pt depth 0pt
                 \kern 0.8pt\hbox{\kern -2pt$\scriptscriptstyle S$}}}}
\def\Rb{\kern 2pt\mathchoice
            {                 \vbox{\hrule width5.5pt height 0.4pt depth 0pt
                 \kern 1.2pt\hbox{\kern -2.5pt$\displaystyle R$}}}
            {                 \vbox{\hrule width5.5pt height 0.4pt depth 0pt
                 \kern 1.2pt\hbox{\kern -2.5pt$\textstyle R$}}}
            {                 \vbox{\hrule width3.5pt height 0.4pt depth 0pt
                 \kern 1.0pt\hbox{\kern -2.2pt$\scriptstyle R$}}}
            {                 \vbox{\hrule width3pt height 0.4pt depth 0pt
                 \kern 0.8pt\hbox{\kern -2.2pt$\scriptscriptstyle R$}}}}
  
\def\pp{{\mathchoice
                                                                                                                                                                                                                                            {              \kern 1pt              \raise 1pt
              \vbox{\hrule width5pt height0.4pt depth0pt
                    \kern -2pt
                    \hbox{\kern 2.3pt
                          \vrule width0.4pt height6pt depth0pt
                          }
                    \kern -2pt
                    \hrule width5pt height0.4pt depth0pt}                    \kern 1pt
           }
            {              \kern 1pt              \raise 1pt
              \vbox{\hrule width4.3pt height0.4pt depth0pt
                    \kern -1.8pt
                    \hbox{\kern 1.95pt
                          \vrule width0.4pt height5.4pt depth0pt
                          }
                    \kern -1.8pt
                    \hrule width4.3pt height0.4pt depth0pt}                    \kern 1pt
            }
            {              \kern 0.5pt              \raise 1pt
              \vbox{\hrule width4.0pt height0.3pt depth0pt
                    \kern -1.9pt                      \hbox{\kern 1.85pt
                          \vrule width0.3pt height5.7pt depth0pt
                          }
                    \kern -1.9pt
                    \hrule width4.0pt height0.3pt depth0pt}                    \kern 0.5pt
            }
            {              \kern 0.5pt              \raise 1pt
              \vbox{\hrule width3.6pt height0.3pt depth0pt
                    \kern -1.5pt
                    \hbox{\kern 1.65pt
                          \vrule width0.3pt height4.5pt depth0pt
                          }
                    \kern -1.5pt
                    \hrule width3.6pt height0.3pt depth0pt}                    \kern 0.5pt            }
        }}
  
\def\mm{{\mathchoice
                                                                                                                                                                                                                                                                            {                             \kern 1pt
               \raise 1pt    \vbox{\hrule width5pt height0.4pt depth0pt
                                  \kern 2pt
                                  \hrule width5pt height0.4pt depth0pt}
                             \kern 1pt}
                       {                            \kern 1pt
               \raise 1pt \vbox{\hrule width4.3pt height0.4pt depth0pt
                                  \kern 1.8pt
                                  \hrule width4.3pt height0.4pt depth0pt}
                             \kern 1pt}
                       {                            \kern 0.5pt
               \raise 1pt
                            \vbox{\hrule width4.0pt height0.3pt depth0pt
                                  \kern 1.9pt
                                  \hrule width4.0pt height0.3pt depth0pt}
                            \kern 1pt}
                       {                           \kern 0.5pt
             \raise 1pt  \vbox{\hrule width3.6pt height0.3pt depth0pt
                                  \kern 1.5pt
                                  \hrule width3.6pt height0.3pt depth0pt}
                           \kern 0.5pt}
                       }}
\def\pd{{\kern0.5pt
                   + \kern-5.05pt \raise5.8pt\hbox{$\textstyle.$}\kern
0.5pt}}
\def\pmd{{\kern0.5pt
                  \pm \kern-5.05pt \raise6.3pt\hbox{$\textstyle.$}\kern1.5pt}}
\def\md{{\mathchoice
   {      {{\kern 1pt - \kern-6.2pt \raise5pt\hbox{$\textstyle.$}\kern 1pt}}}
    {      {{\kern 1pt - \kern-6.2pt \raise5pt\hbox{$\textstyle.$}\kern 1pt}}}
    {      {\kern0.5pt - \kern-5.05pt \raise3.4pt\hbox{$\textstyle.$}\kern0.5pt}}
    {      {\kern0.5pt - \kern-5.05pt \raise3.4pt\hbox{$\textstyle.$}\kern0.5pt}}}}

\begin{document}

\begin{titlepage}
\begin{center}
{\LARGE \bf Super Liouville Black Holes}
\\ \vspace{2cm}
J. Kamnitzer \footnotemark\footnotetext{email:
jkamnitz@student.math.uwaterloo.ca} and 
R.B. Mann \footnotemark\footnotetext{email: 
mann@avatar.uwaterloo.ca} 
\\
\vspace{1cm}
Dept. of Physics,
University of Waterloo
Waterloo, ONT N2L 3G1, Canada\\

\vspace{2cm}
\today\\
\end{center}

\begin{abstract}
We describe in superspace a classical theory of 
of two dimensional $(1,1)$ dilaton supergravity coupled to a
super-Liouville field, and find exact 
super black hole solutions to the field equations that have
non-constant curvature.  We consider the possibility that
a gravitini condensate forms and look at the implications for
the resultant spacetime structure. We find that all such
condensate solutions have a condensate and/or naked 
curvature singularity.
\end{abstract}
\end{titlepage}

\section{Introduction}

The construction of exact solutions of supergravity theories remains a
subject of current interest. In part this is spurred by interest in D-branes 
\cite{deWit}: since D0-branes are massive superparticles, the development of
exact solutions of supergravity theories that contain a superparticle
provides a model system for studying their behaviour. \ More generally, a
study of exact supergravity solutions yields some insight into the behaviour
of supergravity theories beyond their low-energy limit, although the
physical interpretation of the solutions awaits further clarification. \
However the technical issues are formidable and very few exact non-trivial
classical solutions to supergravity theories are known (a list is given in
ref. \cite{sparticle}). Moreover exact superspace supergravity solutions
were non-existent until recently.

Progress was made along these lines by considering a version of
two-di\-men\-sional $(1,1)$ dilaton supergravity \cite{sparticle}. The first
step was taken by constructing a theory of a massive superparticle coupled
to supergravity in a manner that ensured the super stress-energy of the
particle generated the superspace curvature, and the superspace curvature
acted back on the superparticle. \ Exact classical superspace solutions for
the superparticle worldline and the supergravity fields were obtained.
However the resultant superspace consisted of two flat patches joined along
the worldline. \ An extension of this solution for superspaces of constant
curvature was made by adding a cosmological constant to the theory, which is
necessarily negative because of supersymmetry \cite{scosmo}. \ This yielded
the first solutions describing a supersymmetric cosmological black hole,
which is a supersymmetric version of an anti de Sitter black hole in two
spacetime dimensions \cite{ads2dbh}. \ 

Even in two dimensions finding non-trivial solutions (those that cannot by
infinitesimal local supersymmetry transformations be reduced to purely
bosonic solutions) to classical supergravity theories is a difficult task.
The (non)-tri\-vi\-a\-li\-ty of a solution can be determined by the method given in 
\cite{aichel} , which requires solving a differential equation for an
appropriately well-behaved infinitesimal spinor. The virtue of superspace
methods is that this problem is avoided \cite{bible} because a genuine
superspace supergravity solution -- one which satisfies the constraints --
has nonzero torsion beyond that of flat superspace. The torsion is a
supercovariant quantity, and as such its value remains unchanged under a
gauge transformation. Consequently any exact superspace solution with
non-zero torsion must necessarily be non-trivial in this sense.

Here we extend these methods to include $(1,1)$ supergravity coupled to a
super-Liouville field. \ For the supergravity theory we employ a
supersymmetric generalization of the $(1+1)$ dimensional ``$R=T$'' theory,
in which the evolution of the supergravitational fields are determined only
bythe supermatter stress-energy (and vice versa). This stress-energy is
taken to be that of a Liouville field non-minimally coupled to supergravity.
The superdilaton field classically decouples from the evolution of the
supergravity/super-Liouville system, resulting in a two-dimensional theory
that is most closely aligned with $(3+1)$ dimensional supergravity.

We obtain the first super black hole solutions in superspace which have
non-constant curvature. These are generalizations of those found for a
purely bosonic version of this system, in which a Liouville field was
coupled to the bosonic ``$R=T$'' theory \cite{LBH}. The solution space is
potentially just as rich as this case, and we investigate its properties.
For the full superspace solution, we find a broad set of exact solutions 
characterized by three parameters.  We then consider the possibility that
a gravitini condensate forms. We find that, under this circumstance, nearly
all of the exact solutions we obtain develop naked singularities, although
there does exist a single class of new super black hole solutions which 
have reasonable properties.

The outline of our paper is as follows.  
In section 2, we review the bosonic Liouville action and its solution.  
We then give the supersymmetric Liouville action and explain its connection with the 
bosonic case.  
In section 3, we solve the field equations arising from the action and examine some 
features of the solution.  In section 4, we examine the how the causal 
structure of the bosonic 
limit differs if a gravitini condensate forms.  
Finally, in section 5, we look at the energetics of these condensate solutions.

\section{The Supersymmetric action}

We begin by reviewing the solution for the bosonic
Liouville field. The
method differs slightly from that in \cite{LBH} since we choose
conformal coordinates before the variational principle is applied.

In conformal coordinates, the metric takes the form $g_{\pp\mm}=g_{\mm\pp}=\frac{1}{2}e^{2%
\rho }$, $g_{\pp\pp}=g_{\mm\mm}=0$, so that 
\begin{equation}
ds^{2} = e^{2\rho }dx^{\pp}dx^{\mm}  \label{e1}
\end{equation}
where $\sqrt{-g}= \frac{1}{2} e^{2\rho }$, $g^{\pp\mm}=g^{\mm\pp}= 2e^{-2\rho }$ and $%
g^{\pp\pp}=g^{\mm\mm}=0$. From this it is straightforward to derive 
\begin{gather}
R= - 8e^{-2\rho }\partial _{\pp}\partial _{\mm}\rho  \label{e2} \\
(\nabla \phi )^{2}=g^{\mu \nu }\nabla _{\mu }\phi \nabla _{\nu }\phi
= 4e^{-2\rho }\partial _{\pp}\phi \partial _{\mm}\phi  \label{e3}
\end{gather}
Writing the gravitational coupling as $\kappa =4\pi G$, the action for a
Liouville field minimally coupled to $(1+1)$ dimensional gravity is 
\begin{equation}
I=\int d^{2}x\sqrt{-g}\left( \frac{1}{2\kappa }\left( \frac{1}{2}( \nabla \psi
)^{2}+\psi R\right) +b(\nabla \phi )^{2}+\Lambda e^{-2a\phi }+\gamma \phi
R\right)  \label{actb}
\end{equation}
Thus in conformal coordinates: 
\begin{equation}
I=\int d^{2}x \left( \frac{1}{2\kappa }\left( \partial _{\pp}\psi
\partial _{\mm}\psi - 4\psi \partial _{\pp}\partial _{\mm}\rho \right) + 2b\partial
_{\pp}\phi \partial _{\mm}\phi +\frac{\Lambda }{2}e^{2(\rho -a\phi )} - 4\gamma
\phi \partial _{\pp}\partial _{\mm}\rho \right)  \label{actb2}
\end{equation}

The field equations are 
\begin{gather}
4b\partial _{\pp}\partial _{\mm}\phi +a\Lambda e^{2(\rho -a\phi )}+4\gamma
\partial _{\pp}\partial _{\mm}\rho =0  \label{fe1} \\
\frac{2}{\kappa }\partial _{\pp}\partial _{\mm}\psi - \Lambda e^{2(\rho -a\phi
)}+4\gamma \partial _{\pp}\partial _{\mm}\phi =0  \label{fe2} \\
2\partial _{\pp}\partial _{\mm}\psi +4\partial _{\pp}\partial _{\mm}\rho =0
\label{fe3}
\end{gather}

Combining these equations we find: 
\begin{equation}
\partial _{\pp}\partial _{\mm}(\rho -a\phi )= - \frac{\kappa b+2a\gamma \kappa
-a^{2}}{4(\kappa \gamma ^{2}+b)}\Lambda e^{2(\rho -a\phi )}  \label{lbh17}
\end{equation}
which agrees with (17) in \cite{LBH}, except for a difference in sign which is
due to a different choice of conformal coordinates. 

For the supersymmetric version, we start the following action for the
supergravity theory considered in \cite{scosmo}. Choosing the same
normalization coefficients, the action for a super-Liouville field
non-minimally coupled to supergravity is

\begin{equation}
I=4\int d^{2}xd^{2}\theta E^{-1}\left( \frac{1}{2\kappa }\left( \nabla
_{+}\Psi \nabla _{-}\Psi +\Psi R\right) +2b\nabla _{+}\Phi \nabla _{-}\Phi
+Le^{2a\Phi }+\gamma \Phi R\right)   \label{act1}
\end{equation}
where the light-cone coordinates are
$\left( \theta ^{+},\theta ^{-}\right) $ and
$\left( x^{{%
\mathchoice
                                                                                                                                                                                                                                            {              \kern 1pt              \raise 1pt
              \vbox{\hrule width5pt height0.4pt depth0pt
                    \kern -2pt
                    \hbox{\kern 2.3pt
                          \vrule width0.4pt height6pt depth0pt
                          }
                    \kern -2pt
                    \hrule width5pt height0.4pt depth0pt}                    \kern 1pt
           }
            {              \kern 1pt              \raise 1pt
              \vbox{\hrule width4.3pt height0.4pt depth0pt
                    \kern -1.8pt
                    \hbox{\kern 1.95pt
                          \vrule width0.4pt height5.4pt depth0pt
                          }
                    \kern -1.8pt
                    \hrule width4.3pt height0.4pt depth0pt}                    \kern 1pt
            }
            {              \kern 0.5pt              \raise 1pt
              \vbox{\hrule width4.0pt height0.3pt depth0pt
                    \kern -1.9pt                      \hbox{\kern 1.85pt
                          \vrule width0.3pt height5.7pt depth0pt
                          }
                    \kern -1.9pt
                    \hrule width4.0pt height0.3pt depth0pt}                    \kern 0.5pt
            }
            {              \kern 0.5pt              \raise 1pt
              \vbox{\hrule width3.6pt height0.3pt depth0pt
                    \kern -1.5pt
                    \hbox{\kern 1.65pt
                          \vrule width0.3pt height4.5pt depth0pt
                          }
                    \kern -1.5pt
                    \hrule width3.6pt height0.3pt depth0pt}                    \kern 0.5pt            }%
}},x^{{%
\mathchoice
                                                                                                                                                                                                                                                                            {                             \kern 1pt
               \raise 1pt    \vbox{\hrule width5pt height0.4pt depth0pt
                                  \kern 2pt
                                  \hrule width5pt height0.4pt depth0pt}
                             \kern 1pt}
                       {                            \kern 1pt
               \raise 1pt \vbox{\hrule width4.3pt height0.4pt depth0pt
                                  \kern 1.8pt
                                  \hrule width4.3pt height0.4pt depth0pt}
                             \kern 1pt}
                       {                            \kern 0.5pt
               \raise 1pt
                            \vbox{\hrule width4.0pt height0.3pt depth0pt
                                  \kern 1.9pt
                                  \hrule width4.0pt height0.3pt depth0pt}
                            \kern 1pt}
                       {                           \kern 0.5pt
             \raise 1pt  \vbox{\hrule width3.6pt height0.3pt depth0pt
                                  \kern 1.5pt
                                  \hrule width3.6pt height0.3pt depth0pt}
                           \kern 0.5pt}}}\right) =\frac{1}{2}\left( x^{1}\pm
x^{0}\right) $. Here $\Phi $ is
the super-Liouville field, $R$ is the scalar supercurvature, $E=$ sdet$\left[
E_{A}^{\;M}\right] $, the superdeterminant of the vielbein, and $\Psi $ is
the dilaton superfield. The supercovariant derivatives are $\nabla _{\pm }$.  Using 
$ (D_+, D_-) = (\partial_+ + i \theta^+ \partial_\pp, \partial_- + i \theta^- \partial_\mm) $, 
we express them in conformal gauge as 
\begin{align}
\nabla _{+}& =e^{S}(D_{+}+2(D_{+}S)M)  \label{e4} \\
\nabla _{-}& =e^{S}(D_{-}-2(D_{-}S)M)  \label{e5} \\
R& =4e^{S}(D_{-}D_{+}S)  \label{e6} \\
E& =e^{-2S}  \label{e7}
\end{align}
They satisfy the $(1,1)$ supergravity constraints \cite{jim},\cite{rocek} 
\begin{eqnarray}
\left\{ \nabla _{+},\nabla _{+}\right\}  &=&2i\nabla _{{{%
\mathchoice
                                                                                                                                                                                                                                            {              \kern 1pt              \raise 1pt
              \vbox{\hrule width5pt height0.4pt depth0pt
                    \kern -2pt
                    \hbox{\kern 2.3pt
                          \vrule width0.4pt height6pt depth0pt
                          }
                    \kern -2pt
                    \hrule width5pt height0.4pt depth0pt}                    \kern 1pt
           }
            {              \kern 1pt              \raise 1pt
              \vbox{\hrule width4.3pt height0.4pt depth0pt
                    \kern -1.8pt
                    \hbox{\kern 1.95pt
                          \vrule width0.4pt height5.4pt depth0pt
                          }
                    \kern -1.8pt
                    \hrule width4.3pt height0.4pt depth0pt}                    \kern 1pt
            }
            {              \kern 0.5pt              \raise 1pt
              \vbox{\hrule width4.0pt height0.3pt depth0pt
                    \kern -1.9pt                      \hbox{\kern 1.85pt
                          \vrule width0.3pt height5.7pt depth0pt
                          }
                    \kern -1.9pt
                    \hrule width4.0pt height0.3pt depth0pt}                    \kern 0.5pt
            }
            {              \kern 0.5pt              \raise 1pt
              \vbox{\hrule width3.6pt height0.3pt depth0pt
                    \kern -1.5pt
                    \hbox{\kern 1.65pt
                          \vrule width0.3pt height4.5pt depth0pt
                          }
                    \kern -1.5pt
                    \hrule width3.6pt height0.3pt depth0pt}                    \kern 0.5pt            }%
}}},\text{ \ \ \ \ \ \ \ \ \ }\left\{ \nabla _{-},\nabla _{-}\right\}
=2i\nabla _{{{%
\mathchoice
                                                                                                                                                                                                                                                                            {                             \kern 1pt
               \raise 1pt    \vbox{\hrule width5pt height0.4pt depth0pt
                                  \kern 2pt
                                  \hrule width5pt height0.4pt depth0pt}
                             \kern 1pt}
                       {                            \kern 1pt
               \raise 1pt \vbox{\hrule width4.3pt height0.4pt depth0pt
                                  \kern 1.8pt
                                  \hrule width4.3pt height0.4pt depth0pt}
                             \kern 1pt}
                       {                            \kern 0.5pt
               \raise 1pt
                            \vbox{\hrule width4.0pt height0.3pt depth0pt
                                  \kern 1.9pt
                                  \hrule width4.0pt height0.3pt depth0pt}
                            \kern 1pt}
                       {                           \kern 0.5pt
             \raise 1pt  \vbox{\hrule width3.6pt height0.3pt depth0pt
                                  \kern 1.5pt
                                  \hrule width3.6pt height0.3pt depth0pt}
                           \kern 0.5pt}}}}  \label{sgc1} \\
\left\{ \nabla _{+},\nabla _{-}\right\}  &=&RM\text{ \ \ \ \ \ \ \ \ \ \ \ \
\ \ \ \ \ \ \ \ \ \ \ }T_{+{%
\mathchoice
                                                                                                                                                                                                                                            {              \kern 1pt              \raise 1pt
              \vbox{\hrule width5pt height0.4pt depth0pt
                    \kern -2pt
                    \hbox{\kern 2.3pt
                          \vrule width0.4pt height6pt depth0pt
                          }
                    \kern -2pt
                    \hrule width5pt height0.4pt depth0pt}                    \kern 1pt
           }
            {              \kern 1pt              \raise 1pt
              \vbox{\hrule width4.3pt height0.4pt depth0pt
                    \kern -1.8pt
                    \hbox{\kern 1.95pt
                          \vrule width0.4pt height5.4pt depth0pt
                          }
                    \kern -1.8pt
                    \hrule width4.3pt height0.4pt depth0pt}                    \kern 1pt
            }
            {              \kern 0.5pt              \raise 1pt
              \vbox{\hrule width4.0pt height0.3pt depth0pt
                    \kern -1.9pt                      \hbox{\kern 1.85pt
                          \vrule width0.3pt height5.7pt depth0pt
                          }
                    \kern -1.9pt
                    \hrule width4.0pt height0.3pt depth0pt}                    \kern 0.5pt
            }
            {              \kern 0.5pt              \raise 1pt
              \vbox{\hrule width3.6pt height0.3pt depth0pt
                    \kern -1.5pt
                    \hbox{\kern 1.65pt
                          \vrule width0.3pt height4.5pt depth0pt
                          }
                    \kern -1.5pt
                    \hrule width3.6pt height0.3pt depth0pt}                    \kern 0.5pt            }%
}}^{A}=T_{-{{%
\mathchoice
                                                                                                                                                                                                                                                                            {                             \kern 1pt
               \raise 1pt    \vbox{\hrule width5pt height0.4pt depth0pt
                                  \kern 2pt
                                  \hrule width5pt height0.4pt depth0pt}
                             \kern 1pt}
                       {                            \kern 1pt
               \raise 1pt \vbox{\hrule width4.3pt height0.4pt depth0pt
                                  \kern 1.8pt
                                  \hrule width4.3pt height0.4pt depth0pt}
                             \kern 1pt}
                       {                            \kern 0.5pt
               \raise 1pt
                            \vbox{\hrule width4.0pt height0.3pt depth0pt
                                  \kern 1.9pt
                                  \hrule width4.0pt height0.3pt depth0pt}
                            \kern 1pt}
                       {                           \kern 0.5pt
             \raise 1pt  \vbox{\hrule width3.6pt height0.3pt depth0pt
                                  \kern 1.5pt
                                  \hrule width3.6pt height0.3pt depth0pt}
                           \kern 0.5pt}}}}^{A}  \label{sgc2}
\end{eqnarray}
where we have also given the relevant expressions for the vielbein and the
curvature.

Since $\Phi $ and $\Psi $ are scalars, (\ref{act1}) becomes:
\begin{eqnarray}
I &=&4\int d^{2}xd^{2}\theta \left( \frac{1}{2\kappa }\left( D_{+}\Psi
D_{-}\Psi +4\Psi D_{-}D_{+}S\right) +2bD_{+}\Phi D_{-}\Phi \right.   \notag
\\
&&\left. +Le^{2(a\Phi -S)}+\gamma \Phi D_{-}D_{+}S\right)   \label{act1r}
\end{eqnarray}

We next consider the relationship between this action and its bosonic
counterpart in (\ref{actb2}). Setting the fermionic components of the
superfields to zero, we write $S=-\frac{1}{2}\rho +\sigma \theta ^{+}\theta ^{-}$, $%
\Phi =-\frac{1}{2}\phi +B\theta ^{+}\theta ^{-}i$, and $\Psi =-\frac{1}{2}\psi +A\theta
^{+}\theta ^{-}$. Substituting this into (\ref{act1r}) , we find:

\begin{eqnarray}
I &=&4\int d^{2}x\left( \frac{1}{2\kappa }\left( \frac{1}{4}\partial _{\pp%
}\psi \partial _{\mm}\psi -A^{2}-\psi \partial _{\pp}\partial _{\mm}\rho
-4A\sigma \right) \right. \notag \\ 
&+&(2\sigma -2aB)Le^{\rho -a\phi }   
\left. +\frac{b}{2}\partial _{\pp}\phi \partial _{\mm}\phi
-2bB^{2}-4\gamma \sigma B-\gamma \phi \partial _{\pp}\partial _{\mm}\rho
\right)   \label{e8}
\end{eqnarray}
where $\int d^{2}\theta \;\theta ^{+}\theta ^{-}=-1$. \ Varying with respect
to the auxiliary fields $A$ and $B$\ gives: 
\begin{eqnarray}
A &=&-2\sigma   \label{e9a} \\
B &=&-\frac{aLe^{\rho -a\phi }+2\gamma \sigma }{2b}  \label{e9b}
\end{eqnarray}
thereby yielding 
\begin{eqnarray*}
I &=&4\int d^{2}x\left( \frac{1}{2\kappa }\left( \frac{1}{4}\partial _{\pp%
}\psi \partial _{\mm}\psi + 4 \sigma^2  - \psi \partial _{\pp}\partial _{\mm}\rho \right) +
\frac{b}{2}\partial _{\pp}\phi \partial _{\mm}\phi \right.  \\
&&+\left( \frac{a\gamma +b}{b}\right) 2\sigma Le^{\rho -a\phi }+\frac{%
L^{2}a^{2}}{2b}e^{2(\rho -a\phi )}+\frac{2\gamma ^{2}\sigma ^{2}}{b}-\gamma
\phi \partial _{\pp}\partial _{\mm}\rho 
\end{eqnarray*}
upon insertion of (\ref{e9a},\ref{e9b}) into (\ref{e8}). Finally, varying
with respect to the auxiliary field $\sigma $ gives 
\begin{equation*}
\sigma =-\frac{\kappa L(a\gamma +b)e^{\rho -a \phi}}{b+\gamma ^{2}\kappa }
\end{equation*}
which gives us

\begin{equation}
\begin{split}
I &=4\int d^{2}x\left( \frac{1}{2\kappa }\left( \frac{1}{4}\partial _{\pp%
}\psi \partial _{\mm}\psi -\psi \partial _{\pp}\partial _{\mm}\rho \right) +%
\frac{b}{2}\partial _{\pp}\phi \partial _{\mm}\phi  \right. \\
&\quad+ \left.
\frac{a^{2}-2a\gamma
\kappa -b\kappa }{2(b+\gamma ^{2}\kappa) }L^{2}e^{2(\rho -a\phi )}-\gamma \phi
\partial _{\pp}\partial _{\mm}\rho \right)   \label{e10}
\end{split}
\end{equation}
as the bosonic limit of the action (\ref{act1}).

Comparing with (\ref{actb2}), we see that provided 
\begin{equation}
4\frac{a^{2} - 2a\gamma \kappa - b\kappa }{b+\gamma ^{2}\kappa }L^{2}=\Lambda .
\label{e11}
\end{equation}
we recover the bosonic action of \cite{LBH}.  Note that even  in the minimally coupled case ($\gamma = 0$) the
sign of $\Lambda $ depends upon the magnitude $a$\ of the exponential
coupling of the Liouville field.  If $ a = \gamma = 0 $, then the Liouville field decouples and (\ref{e11})
becomes the familiar result that supersymmetry produces only anti de Sitter space \cite{scosmo}.

\section{Solutions to the Field Equations}

Varying the action (\ref{act1r})\ with respect to each of the three superfields
gives the field equations
\begin{gather}
2D_{-}D_{+}\Psi +4D_{-}D_{+}S=0  \label{eq1} \\
2bD_{-}D_{+}\Phi +aLe^{2(a\Phi -S)}+2\gamma D_{-}D_{+}S=0  \label{eq2} \\
\frac{1}{2\kappa }4D_{-}D_{+}\Psi -2Le^{2(a\Phi -S)}+4\gamma D_{-}D_{+}\Phi
=0  \label{eq3}
\end{gather}
where we have used the fact that: 
\begin{equation*}
\int d^2 x d^2 \theta D_-X D_+Y = - \int d^2 x d^2 \theta X D_- D_+ Y
= \int d^2 x d^2 \theta D_- D_+ X Y
\end{equation*}
for commuting superfields $X$, $Y$.
Substituting (\ref{eq1}) into (\ref{eq3}) and combining the resulting
equation with (\ref{eq2}) we find that 
\begin{gather}
D_{-}D_{+}U=\alpha e^{-2U}  \label{pdl1} \\
\text{where}\quad U=S-a\Phi \quad \text{and}\quad \alpha =\frac{%
a^{2}-2a\gamma \kappa -b\kappa }{2(b+\gamma ^{2}\kappa )}L  \label{pdlx}
\end{gather}

Now (\ref{pdl1}) is the super-Liouville equation and its general solution was given in 
\cite{arvis}.  We have: 
\begin{equation}
U=-\frac{1}{2}\ln \left( \frac{-iD_{+}F_{+}D_{-}F_{-}}{2\alpha (F_{\pp}-F_{\mm}-iF_{+}F_{-})}\right)   
\label{pdl2}
\end{equation}
where we have the constraints: $D_{-}F_{+}=D_{+}F_{-}=0$ and $D_{+}F_{\pp%
}=iF_{+}D_{+}F_{+}$, $D_{-}F_{\mm}=iF_{-}D_{-}F_{-}$. These equations are
solved by: 
\begin{align}
F_{\pp}& =f_{\pp}\pm i\theta ^{+}\lambda ^{+}\sqrt{\partial _{\pp}f_{\pp}}
\label{Fpp} \\
F_{+}& =\pm \sqrt{\partial _{\pp}f_{\pp}}\left( 1+\lambda ^{+}\frac{%
i\partial _{\pp}\lambda ^{+}}{2\partial _{\pp}f_{\pp}}\right) \theta
^{+}+\lambda ^{+}  \label{Fp} \\
F_{\mm}& =f_{\mm}\pm i\theta ^{-}\lambda ^{-}\sqrt{\partial _{\mm}f_{\mm}}
\label{Fmm} \\
F_{-}& =\pm \sqrt{\partial _{\mm}f_{\mm}}\left( 1+\lambda ^{-}\frac{%
i\partial _{\mm}\lambda ^{-}}{2\partial _{\mm}f_{\mm}}\right) \theta
^{-}+\lambda ^{-}  \label{Fm}
\end{align}
where $f_{\pp}=f_{\pp}(x_{\pp})$ and $f_{\mm}=f_{\mm}(x_{\mm})$ are
commuting and $\lambda _{+}=\lambda _{+}(x_{\pp})$ and $\lambda _{-}=\lambda
_{-}(x_{\mm})$ are anti-commuting.

From (\ref{eq1}), (\ref{eq2}) and (\ref{eq3}), we also find that: 
\begin{equation}
D_{-}D_{+}\Phi =\frac{a-\gamma \kappa }{b\kappa +a\gamma \kappa }D_{-}D_{+}S
\label{phi-S}
\end{equation}
and so
\begin{equation}
\Phi =\frac{a-\gamma \kappa }{b\kappa +a\gamma \kappa }S+\frac{1}{a}H
\label{eqph}
\end{equation}
where $H$ is the general solution of the homogeneous equation, $D_{-}D_{+}H=0
$. Its superfield expansion is 
\begin{equation}
H=H_{+}+H_{-}=h_{\pp}+\nu ^{+}\theta ^{+}+h_{\mm}+\nu ^{-}\theta ^{-}
\label{eqH}
\end{equation}
where $H_{\pm }$ are arbitrary functions of the superpspace coordinates,
with\  $h_{\pp}=h_{\pp}(x_{\pp})$ and $h_{\mm}=h_{\mm}(x_{\mm})$ are
commuting and $\nu _{+}=\nu _{+}(x_{\pp})$ and $\nu _{-}=\nu _{-}(x_{\mm})$
are anti-commuting.

Combining (\ref{pdlx}) and (\ref{eqph}), we find: 
\begin{equation}
S-a\left( \frac{a-\gamma \kappa }{b\kappa +a\gamma \kappa }S+\frac{1}{a}%
H\right) =U  \label{SHUeq}
\end{equation}
and hence that 
\begin{equation}
S=\beta U+H\quad \text{where}\quad \beta =\frac{\kappa (b+a\gamma )}{b\kappa
+2a\gamma \kappa -a^{2}}  \label{ssolve}
\end{equation}
Combining (\ref{ssolve}) with (\ref{eqH}) and (\ref{pdl2}) gives us the
complete general solution for $S$ as desired. 

We now make a change of variables to put the solution for $ S $ into a
form which matches the solutions of \cite{LBH}.
We choose $ H = 0 $ to simplify our solution
and let $ G_\pp = -\sigma^2/F_\pp $, where $ \sigma $ is a constant which we will set
later.  Now if we write $ G_\pp = g_\pp + i \theta^+ \mu^+ \sqrt{\partial_\pp g_\pp}$, 
then equating components gives $ g_\pp = -\sigma^2/f_\pp$ and $ \mu^+ = \lambda^+ \sigma /f_\pp $. 
We also set $ G_+ = \sqrt{\partial _{\pp}g_{\pp}}\left( 1+\mu ^{+}\frac{
i\partial _{\pp}\mu ^{+}}{2\partial _{\pp}g_{\pp}}\right) \theta
^{+}+\mu ^{+} $ for convenience.   

Substituting these component expressions into
(\ref{ssolve}) gives:
\begin{equation}
S = -\frac{\beta}{2} \ln \left( \frac{ -i 
\bigl( D_+ G_+ - i \theta^+ \mu^+ \frac{ \partial_\pp g_\pp}{ g_\pp } \bigr)
  D_- F_-}{2 \alpha \left(
\sigma -  i \theta^+ \mu^+ \frac{\sigma \sqrt{\partial_\pp g_\pp}}{g_\pp} + f_\mm \frac{g_\pp}{\sigma} +
i\theta^- \lambda^- \frac{g_\pp\sqrt{\partial_\mm f_{\mm}} }{ \sigma} - i G_+ F_- \right)}
\right) 
\label{ssolve2}
\end{equation}

To make things a bit simpler, we choose coordinates so that $%
D_{+}G_{+}=1= -iD_{-}F_{-}$.  This
gives us 
$g_{\pp}=x^{\pp}$, $\mu ^{+}=\lambda _{0}^{+}$ and 
$f_{\mm}=-x^{\mm}$, and $\lambda
^{-}=i\lambda _{0}^{-}$, where $\lambda _{0}^{\pm}$ are anticommuting constants.  
The $i$ is taken for convenience.
Substituting into (\ref{ssolve2}) and performing the division, produces 

\begin{equation}
\begin{split}
S &= \frac{\beta }{2}\ln \Bigl( 2\alpha \Bigl( \sigma - \frac{x^{\pp}x^{\mm}}{\sigma} - i \left( \theta ^{+}\lambda
_{0}^{+}\frac{x^\mm}{\sigma} + \theta ^{-}\lambda _{0}^{-}\frac{x^\pp}{\sigma} \right) \\
& \quad + (\theta ^{+}+\lambda _{0}^{+})(\theta
^{-}+\lambda _{0}^{-}) - \theta^{+} \theta^- \lambda_0^+ \lambda_0^- \frac{1}{\sigma} \Bigr) \Bigr)   \label{S-expression}
\end{split}
\end{equation}
which yields the metric and gravitini fields for the curved superspace induced by a super-Liouville
field.

We close this section by expanding $e^{S}$ to determine the functional form
of the metric and gravitino fields. Following the procedure given in \cite{sparticle},
\cite{rocek} we choose a Wess-Zumino gauge such that the
superconformal gauge ($E_{\pm }=e^{S}D_{\pm }$) is compatible with the
ordinary $x$-space conformal gauge. This gives 
\begin{equation}
e^{S}=\epsilon ^{-1/4}+\frac{1}{2}i\left( \theta ^{+}\psi _{\pp}^{+}-\theta
^{-}\psi _{\mm}^{-}\right) +\frac{1}{4}\theta ^{+}\theta ^{-}\epsilon
^{1/4}\left( iA+\psi _{\mm}^{-}\psi _{\pp}^{+}\right)   \label{Sexpand}
\end{equation}
where $\epsilon =\det (\epsilon _{a}^{m})=\det (e^{\rho }\delta
_{a}^{m})=e^{2\rho }$. 

Now we expand (\ref{ssolve2}) to obtain: 
\begin{align}
e^S &= \Bigl( 2\alpha \Bigl( \sigma - \frac{x^{\pp}x^{\mm}}{\sigma} - i \left( \theta ^{+}\lambda
_{0}^{+}\frac{x^\mm}{\sigma} + \theta ^{-}\lambda _{0}^{-}\frac{x^\pp}{\sigma} \right) + (\theta ^{+}+\lambda _{0}^{+})(\theta
^{-}+\lambda _{0}^{-}) \notag \\ 
& \quad - \frac{ \theta^{+} \theta^- \lambda^+ \lambda^- }{\sigma} \Bigr) \Bigr)^{\beta /2} \notag \\ 
& = \Bigl( 2\alpha \Bigl( \sigma - \frac{x^{\pp}x^{\mm}}{\sigma} \Bigr) \Bigr) ^{\beta /2}
\left[ 1- \Bigl( i \left( \theta^+ \lambda_0^+ \frac{ x^\mm}{\sigma} + \theta^- \lambda_0^- \frac{x^\pp}{\sigma} \right) - \frac{ \theta^+ \theta^- \lambda_0^+ \lambda_0^- }{\sigma} \notag \right. \\
&\left. \quad + (\theta^+ + \lambda_0^+)(\theta^- + \lambda_0^-) \Bigr)
\frac{ \beta}{ 2 (  \sigma - \frac{x^\pp x^\mm}{\sigma} ) } 
+ \theta^+ \theta^- \lambda_0^+ \lambda_0^- \frac{ \beta (\beta - 2)}{ (2 (\sigma - \frac{x^\pp x^\mm}{\sigma}) )^2 } 
\right]
\label{expS-eqn}
\end{align}

Comparing this with (\ref{Sexpand})  gives: 
\begin{align}
&e^{-\rho /2} =\left( 2\alpha  (\sigma - \frac{x^{\pp}x^{\mm}}{\sigma})\right)^{\beta /2} \left( 1 +
\frac{ \beta \lambda _{0}^{+}\lambda _{0}^{-}}{2(\sigma - \frac{x^{\pp}x^{\mm}}{\sigma})}\right) 
\label{fesol1} \\
&\psi _{\pp}^{+} =
 -\beta ( 2\alpha )^{\beta/2} \left( \sigma - \frac{x^{\pp}x^{\mm}}{\sigma} \right) ^{\beta /2 -1} \left( \lambda _{0}^{+} \frac{x^\mm}{ \sigma} + i \lambda _{0}^{-} \right)  
\label{fesol2} \\
&\psi _{\mm}^{-} = \beta ( 2\alpha )^{\beta/2} \left( \sigma - \frac{x^{\pp}x^{\mm}}{\sigma}
\right) ^{\beta /2 -1} \left( \lambda _{0}^{-} \frac{x^\pp }{ \sigma} - i \lambda _{0}^{+} \right)  
\label{fesol3} \\
&A = -2i \beta (2\alpha )^\beta \left( \sigma - \frac{x^{\pp}x^{\mm}}{\sigma} \right) ^{\beta -2} \left( \sigma - \frac{x^{\pp}x^{\mm}}{\sigma}+ (\beta - 1) \lambda _{0}^{+} \lambda _{0}^{-} \right)  
\label{fesol4}
\end{align}

To complete the solution, we give the super-Dilaton and super-Liouville fields, $ \Psi $ and $ \Phi $ respectively. 
These are derived from by substituting $ S $ into (\ref{eq1}) and (\ref{eqph}) respectively.
\begin{align}
&\Psi = -\beta \ln \Bigl( 2\alpha \Bigl( \sigma - \frac{x^{\pp}x^{\mm}}{\sigma} - i \left( \theta ^{+}\lambda
_{0}^{+}\frac{x^\mm}{\sigma} + \theta ^{-}\lambda _{0}^{-}\frac{x^\pp}{\sigma} \right) \notag \\
&\quad \quad + (\theta ^{+}+\lambda _{0}^{+})(\theta
^{-}+\lambda _{0}^{-}) - \theta^{+} \theta^- \lambda_0^+ \lambda_0^- \frac{1}{\sigma} \Bigr) \Bigr) + J \label{psisolve} \\
&\Phi = -\frac{\beta -1}{2a} \ln \Bigl( 2\alpha \Bigl( \sigma - \frac{x^{\pp}x^{\mm}}{\sigma} - i \left( \theta ^{+}\lambda
_{0}^{+}\frac{x^\mm}{\sigma} + \theta ^{-}\lambda _{0}^{-}\frac{x^\pp}{\sigma} \right) \notag \\ 
&\quad \quad+ (\theta ^{+}+\lambda _{0}^{+})(\theta
^{-}+\lambda _{0}^{-}) - \theta^{+} \theta^- \lambda_0^+ \lambda_0^- \frac{1}{\sigma} \Bigr) \Bigr) 
\label{phisolve}
\end{align}
where $ J $ is a solution to $ D_-D_+ J = 0 $ (see (\ref{eqH})).
 
Several structural features of the superspace are already apparent from the
preceding relations.  If $ b = -a\gamma $ then $\beta =0$.  This yields a
flat superspace \cite{sparticle} and $ \Phi $  obeys the flat super-Liouville equation.  If $ \gamma \kappa 
= a $ then $\beta =1$.  This corresponds to the anti de Sitter
superspace \cite{scosmo} and $ \Phi $ is a free scalar superfield in a superspace of 
constant negative curvature.
In the minimally coupled
theory these situations
respectively correspond to a constant scalar superfield with no kinetic
energy term and a free scalar superfield with no coupling.

The solutions (\ref{fesol1}--\ref{fesol4}) are not
gauge transforms of a purely gravitational solution in  dilaton gravity, but
are indeed non-trivial as they have non-zero torsion.  They furnish us with a
rich 3-parameter set of supersymmetric black hole solutions that are extensions of those 
found in ref. \cite{LBH}.  These
solutions have non-constant superspace curvature everywhere except for the
special cases $\beta =0,2$ mentioned above.

There are a number of physical interpretations of these solutions one could make.
One possibility is interpret them as relevant only when the gravitini vanish. In
this case our 3-parameter set of solutions reduces to those found in the bosonic
case  \cite{LBH}.  However since the correction to the
conformal factor of the metric is proportional to ${\psi_\pp}^+ {\psi_\mm}^-$,
a condensate of gravitini will modify the structure of spacetime.  In the next
section we consider this possibility.

\section{Bosonic Condensate}

We now examine the possibility that in the bosonic limit, the gravitino
and gravitini fields form a condensate.  That is relative to some local
vacuum $ \langle  \lambda_0^+ \rangle = 0 = \langle \lambda_0^- \rangle $ 
but $ \langle \lambda_0^+ \lambda_0^- \rangle = c \ne 0 $ where the phases of
$\lambda_0^+$ and $\lambda_0^- $ are chosen so that $c$
is some real commuting constant.  In this case, the gravitino and gravitini fields vanish 
($ \psi_\pp^+ = 0 = \psi_\mm^- $) but the gravitini condensate becomes 
spacetime dependent:
\begin{equation}\label{gravcond}
\langle \psi _{\pp}^{+} \psi _{\mm}^{-}  \rangle 
= \frac{c\beta^2}{\sigma} (2\alpha )^\beta 
\left( \sigma - \frac{x^{\pp}x^{\mm}}{\sigma} \right) ^{\beta -1}
\end{equation}
and we also obtain
$ ds^2 = \langle e^{2\rho} \rangle dx^{\pp} dx^{\mm} $, which becomes
\begin{equation}
ds^2 =  \left( 2\alpha  (\sigma - \frac{x^{\pp}x^{\mm}}{\sigma})\right)^{-2\beta} 
\left( 1 -\frac{2  \beta c}{\sigma - \frac{x^{\pp}x^{\mm}}{\sigma}} \right) 
dx^{\pp} dx^{\mm} \label{gmet1}
\end{equation}
for the bosonic space-time metric. 

We choose $ \sigma = - 2 \alpha / K^2 $ and do a coordinate transformation 
of $ x^\mm = - x^\mm $ in order to connect with \cite{LBH}.
This gives us: 
\begin{equation}
ds^2 = - \left( \frac{- 4 \alpha^2}{K^2} - K^2 {x^{\pp}x^{\mm}} \right)^{-2\beta} 
\left( 1 + \frac{2 \beta c}{\frac{2\alpha}{K^2} 
+ \frac{K^2}{2 \alpha} x^{\pp}x^{\mm} } \right) dx^{\pp} dx^{\mm} \label{gmet2}
\end{equation}
Note that in the $ c = 0 $ limit we recover the metric induced by a bosonic Liouville
field (eq. (23) of \cite{LBH}) with $ -4 \alpha^2 = A $.  
This is consistent with (\ref{e11}).  We see that supersymmetry forces A to be negative.

We next consider how the gravitini condensate affects 
the causal structure of the spacetime. The curvature scalar is 
\begin{align}
R &=  -16 \beta\alpha K^{4(\beta-1)}\frac{(-(4\alpha^2+K^4 x^{\pp}x^{\mm}))^{2\beta-1}}
{(4\alpha^2+4\beta\alpha c K^2+K^4 x^{\pp}x^{\mm})^3}  \nonumber \\
& \qquad \times\left( K^8(2\alpha-c K^2)(x^{\pp}x^{\mm})^2 
+ 16\alpha^2 K^4(\alpha+\beta cK^2)(x^{\pp}x^{\mm}) 
+ 16 c \alpha^4 (cK^2+2\alpha+4\beta cK^2)
\right)
\label{Ricxx}
\end{align}
and we see that its singularity structure depends upon the values of the parameters.
For all values of $\beta$, there is a singularity at 
$x^{\pp}x^{\mm}=-4\frac{\alpha^2}{K^4}(1+\frac{\beta c K^2}{\alpha})$. 
For $\beta > 1$  the curvature scalar $R$ diverges at $x^{\pp}x^{\mm} \to \pm\infty$,
and for $\beta<1/2$, it diverges at $x^{\pp}x^{\mm} \to -4\frac{\alpha^2}{K^4}$.
The volume element of the metric either diverges or vanishes at $x^{\pp}x^{\mm}=-4\frac{\alpha^2}{K^4}$ 
depending on the sign of $\beta$, and the (\ref{gmet2}) is therefore only valid 
$x^{\pp}x^{\mm}<-4\frac{\alpha^2}{K^4}$.   This region will have no curvature singularities
if $1/2 < \beta < 1$ and $\frac{\beta c K^2}{\alpha}<0$.

The condensate (\ref{gravcond}) transforms as a scalar density, and so we write
\begin{equation}\label{vphi}
\langle \psi^{+} \psi^{-}  \rangle = e^{-i\theta}\sqrt{-g}\varphi 
\end{equation}
where $\varphi$ transforms as a scalar, and the phase
$\theta$ may be chosen so that 
$\varphi$ is always real. Its singular behaviour will be
co-ordinate invariant.  We obtain
\begin{equation}\label{vphixx}
\varphi = -e^{i\theta}{K^4 c\beta^2}(2\alpha )^{2\beta -1}
\frac{ \left( \frac{- 4 \alpha^2}{K^2} - K^2 {x^{\pp}x^{\mm}} \right)^{3\beta} }
{(4\alpha^2+4\beta\alpha c K^2+K^4 x^{\pp}x^{\mm})}
\end{equation}
and so,  as with the curvature, there is an 
additional divergence at 
$x^{\pp}x^{\mm}=-4\frac{\alpha^2}{K^4}(1+\frac{\beta c K^2}{\alpha})$ for all
values of $\beta$. We also see that for $\beta < 0$  the condensate $\varphi$
diverges at $x^{\pp}x^{\mm} \to -4\frac{\alpha^2}{K^4}$, whereas if $\beta>1/3$ it
diverges at $x^{\pp}x^{\mm} \to \pm\infty$. As with the curvature, there is an 
additional divergence at 
$x^{\pp}x^{\mm}=-4\frac{\alpha^2}{K^4}(1+\frac{\beta c K^2}{\alpha})$ for all
values of $\beta$. If $\frac{\beta c K^2}{\alpha}<0$, then the range of coordinates
is restricted by $x^{\pp}x^{\mm}<-4\frac{\alpha^2}{K^4}$, and
this region will have no invariant condensate singularities
if $0 < \beta \leq 1/3$. There is no nonzero value of $\beta$ for which the spacetime is
free of condensate and coordinate singularities. 

If $2\beta$ is an integer then (\ref{gmet2}) is valid for all
$x^{\pp}x^{\mm}\neq -4\frac{\alpha^2}{K^4}$, 
with $x^{\pp}x^{\mm} = -4\frac{\alpha^2}{K^4}$
forming an asymptotic region of either of two distinct spacetimes.  As noted
previously, if $\beta=0$ the curvature and condensate vanish and
the spacetime is flat. If $\beta=\pm 1/2$, the only singularity in the curvature is at 
$x^{\pp}x^{\mm}=-4\frac{\alpha^2}{K^4}(1+ \beta  m)$, where $m \equiv
\frac{c K^2}{\alpha}$ is a dimensionless quantity which parametrizes the strength
of the condensate.

Proceeding further, we switch to more familiar $(r,t)$ coordinates 
via the transformation 
\begin{equation}\label{xxtfm}
x^{\pp}x^{\mm} = -\frac{4\alpha^2}{K^4}(u(r)+1) 
\qquad   \frac{x^{\pp}}{x^{\mm}} = \exp(2ht)
\end{equation}
which yields
\begin{equation}
ds^2 = \frac{u - \beta m}{u^{2 \beta + 1}} 
\left(\frac{(u^\prime dr)^2}{4 h^2 (u+1)} - (u+1) dt^2 \right)
\label{gmet2a}
\end{equation}
in place of (\ref{gmet2}), where prime denotes the derivative with respect to $r$ and
$h = (4 \alpha^2)^{\beta - 1/2} K^{2-2\beta }$, which we take to be positive

The range of the function $u(r)$ must be chosen to ensure that the metric is real
and of the appropriate signature. If $2\beta$ is not an integer, 
then we must have $u(r)>0$, whereas otherwise
the reality requirement is trivially met for any choice of sign for $u(r)$, with
$u(r)=0$ corresponding to an asymptotic region of spacetime.

By choosing $u(r)$ so that $ u^\prime = 2h u^{2\beta}$, we 
obtain the metric in the form
\begin{equation}
ds^2 = \left( \frac{dr^2}{2Mr \pm (2Mr)^{2\beta/(2\beta-1)}} 
-\left[2Mr \pm (2Mr)^{2\beta/(2\beta-1)}\right] dt^2) \right)
\left( 1 \mp  \beta m (2Mr)^{1/(2\beta-1)}) \right) 
\label{gmet2b}
\end{equation}
provided $\beta\neq 1/2$ where $M=h(1-2\beta)$, and 
both signs apply if $\beta$ is some other odd half-integer (otherwise
only the plus sign applies). For $\beta=1/2$, 
\begin{equation}
ds^2 =  \left( \frac{dr^2}{1 \pm e^{-2hr}} 
-\left[1 \pm e^{-2hr}\right] dt^2) \right)
\left( 1 \mp m e^{-2hr}) \right) 
\label{gmet2c}
\end{equation}
These metrics are the same as those given in ref. \cite{LBH} (with $A<0$),
apart from a conformal factor due to the gravitini condensate.   For $\beta \leq 1/2$
the behaviour of the spacetime at infinity is not modified relative to these
pure bosonic cases.

We turn now to consider specific cases, moving to 
Schwarzschild-type coordinates by selecting $u$ so that
\begin{equation}\label{ueqn}
( 1 - \frac{\beta m}{u}) u^\prime = 2 h u^{2\beta}
\end{equation}
Henceforth the function $u$ will be chosen to satisfy this equation, whose solution is
\begin{equation}\label{usol1}
{2h r} = \frac{1-u^{1-2\beta }}{2\beta -1} + \frac{m u^{-2 \beta}}{2}  
\end{equation}
where the constant of integration has been chosen so that the $\beta\to \frac{1}{2}$ limit
can be straightforwardly taken:
\begin{equation}\label{usol2}
{2hr}  = \frac{u \ln(u^2 /k) + m}{2u}  
\end{equation}
where the constant $k$ is arbitrary. In the $(r,t)$ coordinates,
the gravitini condensate (\ref{vphixx}) becomes
\begin{equation}\label{gravcond2}
\varphi  = - e^{i\theta} \frac{m\beta h^3(2\alpha)^{2\beta}}{2K^4} 
\frac{u^{3\beta}}{u-\beta m}
\end{equation}
and the curvature scalar is
\begin{equation}\label{Rscal2}
R  = 4 h^2\frac{u^{2\beta-1}}{(u-\beta m)^3}\left[(m-2)u^2+2m(2\beta+1)u+\beta m^2 (2\beta+1)\right]
\end{equation}

The function $r(u)$ in (\ref{usol1}) has a single minimum at $u= \beta m$ for $u>0$. 
If $\beta < 1/2$ it
will diverge as $u\to \infty$, whereas if $\beta > 1/2$ it will asymptote to a constant.
Inverting (\ref{usol1}) we can obtain $ u(r) $ on any interval with 
$ r^\prime (u) \ne 0 $,  i.e. any interval not containing $ 0, m\beta $. 
If $2 \beta$ is not an integer then we are restricted to $ u > 0 $, since 
we are raising it to a fractional power.  Hence if $m\beta  > 0 $ then we get two 
possible spacetimes $m\beta \le u < \infty $ and $ 0 \le u \le m\beta $, and there
is a curvature and condensate singularity at $u=m\beta$ in each.  On the other hand,
if $ m\beta \le 0 $ then there will be only one spacetime, defined for $ 0 \le u < \infty $.
However  freedom from condensate singularities forces $0<\beta<1/3$, whereas freedom from
coordinate singularities forces $1>\beta>1/2$, and so there is no singularity-free region
of parameter space as noted above.

If $ 2 \beta$ is an integer, more possibilities emerge since
the  spacetime will be defined for $ u < 0 $ as well.  The second derivative of 
$r(u)$ is $(\beta m)^{-(1+2\beta)}$, and so is always positive if $2\beta$ is an
odd integer, yielding a minimum at $u= \beta m$. However if $2\beta$ is an
even integer, $r(u)$ will have a maximum at $u= \beta m$ if $\beta m < 0$, and a
minimum at $u= \beta m$ if $\beta m > 0$.   In either case we have three possible
spacetimes ($0> \beta m > u$, $0 > u \geq \beta m $, and $u>0$) for $\beta m$ negative
and three possible spacetimes ($u > \beta m > 0$, $ \beta m \geq u > 0$, and $0 > u$).
The metric has the form
\begin{equation}\label{schmet}
ds =  - f(r)dt^2+\frac{dr^2}{f(r)}  
\end{equation}
where
\begin{equation}\label{schmetf}
{f(r)}  = \frac{(u-\beta m)(u+1)}{u^{2\beta+1}}
\end{equation}
with $u(r)$ implicitly defined in (\ref{usol1},\ref{usol2}) above.  There is now the 
possibility of placing the singularity behind a horizon at $u=-1$.  

We now need only demand that the curvature and
condensate be finite outside the black hole (ie for $u>-1$).    
Implicitly plotting the function (\ref{schmetf}) 
allows us to find the causal structure of the spacetimes.  
We construct the Penrose diagrams using the method outlined in \cite{Strobl},
where we represent a horizon with a dashed line, a non-singular
boundary with a thin black line is, and a singularity with a solid black line.

Searching through parameter space, only for $\beta=-1/2$ do we find a spacetime
whose curvature singularity is shielded by an event horizon. However the condensate
singularity diverges at $u=0$, outside of the event horizon where $u=-1$.

We close this section by computing the quasilocal energy of the solutions we obtain. 
Using methods of \cite{quasi} we find from
(\ref{psisolve}) that the dilaton field is
\begin{align} \Psi &= -\beta \ln \left( \frac{4\alpha^2}{K^2} \right) - \beta \ln (u + \frac{m}{2 \beta}) 
\\ 
\Rightarrow \frac{ d \Psi}{ dr } &= \frac{ d \Psi}{ du } \frac{ d u}{ dr } 
= \frac{-4h \beta u^{2 \beta + 1}}{(2 u + m) (u-\beta m)} 
\end{align} 
taking the gravitino-gravitini condensate into account.  The quasilocal energy has the
generic form \cite{quasi}
\begin{equation} 
E = 2(N_{ref} - N) \frac{d \psi}{dr} \label{enfor} 
\end{equation}
where $ N(r) = \sqrt{f(r)}$.  

The choice of reference spacetime is somewhat problematic for the class of solutions we consider
due to their varying asymptotic properties.  For simplicity we set $ N_{ref} = 0 $ (a reference value) 
obtaining
\begin{equation}\label{quasiE}
E = \frac{ 8h \beta u^{\beta + 1/2} }{(2u + m) } \sqrt{ \frac{u+1}{u-\beta m}}
\end{equation} 
For large $u$, which is large $r$, we find that the quasilocal energy either diverges for
$\beta > 1/2$, or else vanishes for $\beta < 1/2$.  To extract further physical meaning from
these solutions will require a more judicious choice of $N_{ref}$.  However 
if $ \beta = 1/2 $, (\ref{quasiE}) becomes
\begin{equation}
E = 2K 
\end{equation}
in the large-$r$ limit, 
indicating that the parameter $K$ corresponds to a mass parameter.

\section{Summary}

We have obtained a new class of exact solutions in $(1+1)$ dimensional
supergravity coupled to a super-Liouville field.  This class of solutions
depends on the three parameters $\beta$, $K$ and $m$.  The first of these
is a function purely of the coupling parameters of the super-Liouville
field to supergravity.  The second of these is a constant of integration,
which is related to the overall mass-energy of the solutions.  The
third is a parameter related to the strength of the gravitini condensate.
Overall, it is the combination of the constants $ \beta $ and $ m $ which 
determine the causal structure of the spacetime.  
If the condensate vanishes then the solutions we obtain reduce to those of
ref. \cite{LBH}.

To our knowledge these solutions are the first exact superspace solutions
of supermatter coupled to supergravity which have non-constant superspace
curvature everywhere.  Their interpretation remains to be clarified, and we have
made a first attempt here by considering the possibility that the gravitini
might form a condensate.  If this does indeed happen in the manner we
consider, then the structure of spacetime is considerably modified relative
to the bosonic solutions of ref. \cite{LBH}.  For all ranges of the parameters,
the solutions we obtain typically have either a naked curvature singularity
or a condensate singularity outside of the event horizon.  
If the condensate singularity could
somehow be eliminated (perhaps by quantum effects), then the spacetime with
$\beta=-1/2$, $m\neq 2$) would yield a solution with a curvature singularity cloaked
by an event horizon. This would appear to be the only `physically' acceptable 
super Liouville black hole in the presence of a gravitini condensate.

\bigskip

\noindent {\bf Acknowledgments}

This research was supported the Natural Sciences and Engineering Research 
Council of Canada. R.B.M. would like to thank M. Grisaru and M.E. Knutt
for discussions.

\end{document}